# Measuring Teachers' Visual Expertise Using the Gaze Relational Index Based on Real-world Eye-tracking Data and Varying Velocity Thresholds

Christian Kosel*[1], Angelina Mooseder*[2], Tina Seidel[1], Jürgen Pfeffer[2]

*These authors contributed equally

[1] Friedl Schöller Endowed Chair for Educational Psychology (Seidel),

School of Social Science and Technology, Technical University Munich, Germany

[2] Computational Social Science and Big Data (Pfeffer),

School of Social Science and Technology, Technical University Munich, Germany

{christian.kosel, angelina.mooseder, tina.seidel, juergen.pfeffer}@tum.de

## Abstract

This article adds to the understanding of teachers' visual expertise by measuring visual information processing in real-world classrooms (mobile eye-tracking) with the newly introduced Gaze Relational Index (GRI) metric, which is defined as the ratio of mean fixation duration to mean fixation number. In addition, the aim was to provide a methodological contribution to future research by showing to what extent the selected configurations (i.e. varying velocity thresholds and fixation merging) of the eye movement event detection algorithm for detecting fixations and saccades influence the results of eye-tracking studies. Our study leads to two important take-home messages: First, by following a novice-expert paradigm (2 novice teachers & 2 experienced teachers), we found that the GRI can serve as a sensitive measure of visual expertise. As hypothesized, experienced teachers' GRI was lower, suggesting that their more fine-graded organization of domain-specific knowledge allows them to fixate more rapidly and frequently in the classroom. Second, we found that the selected velocity threshold parameter alter and, in the worst case, bias the results of an eye-tracking study. Therefore, in the interest of further generalizability of the results within visual expertise research, we emphasize that it is highly important to report configurations that are relevant for the identification of eye movements.

## 1 Introduction

Human visual expertise in vision-intensive domains such as medicine (Gegenfurtner & Seppänen, 2013), sports (Aglioti et al., 2008), driving (Lappi et al., 2017), and aviation (Peißl et al., 2019) reflects complex cognitive and visual processing that evolves in domain experts through deliberate and consistent practice over a long period (Gegenfurtner et al., 2011). Domain experts are more skilled in the perception, interpretation, and evaluation of domain-specific visual information (Gegenfurtner, 2020). In recent years, there has been a growing interest in the visual expertise of teachers because they perceive and interpret a large amount of dynamic visual information to effectively manage the complexity of a classroom full of students. A rich body of studies has revealed that experienced and novice teachers differ markedly in their visual processing (see meta-analysis/reviews: Gegenfurtner et al., 2011; Grub et al., 2020). Besides studies based on verbal reports and think-a-loud protocols that were collected after a visual stimulus (i.e., video vignette, photograph; van Es & Sherin, 2010) was shown to the participating teachers, other studies explored teachers' visual expertise based on fine-graded data that were collected with eye-tracking devices (Kosel, Holzberger, & Seidel, 2021; McIntyre & Foulsham, 2018). Eye-tracking is an effective method to explore where, how often and how long teachers direct their visual attention (Holmqvist et al., 2015). In visual expertise research across various disciplines including teaching, two eye-tracking parameters are found to be commonly sensitive to expertise – the *number of fixations* and the *average fixation duration* (Gegenfurtner et al., 2011; Grub et al., 2020). Generally, it was found in several studies that domain experts, including experienced teachers, have more but shorter fixations, whereas domain novices have fewer but longer fixations (Grub et al., 2020; Wolff, et al., 2016). These findings underline that the processing of visual information not only varies across individuals but also across different levels of expertise (Gegenfurtner et al., 2011). Until now, however, research on visual expertise has lacked a common and single metric to capture and contrast the visual expertise of domain experts and novices. Gegenfurtner and colleagues (2020) fill this gap by introducing a novel eye-tracking metric indicative of expert visual processing, called the Gaze Relational Index (GRI). The GRI is defined as the ratio of mean fixation duration to mean fixation count. As far as we know, GRI has only been applied in two studies that used data from laboratory and stationary eye movement recordings. Gegenfurtner et al. (2020) investigated the GRI regarding 3D dynamic medical visualizations in diagnostic radiology (Gegenfurtner et al., 2020), and Grub, Biermann, Lewalter & Brünken (2022) focused on experienced and novice teachers' gazes using video vignettes (Grub et al., 2022). Studies found marginal expertise differences (Gegenfurtner et al., 2020) or no differences (Grub et al., 2022). The present study aims to go beyond laboratory-collected eye movement data (on-action) and implement the GRI into research on teachers' visual expertise by calculating the GRI for experienced and novice teachers' real-world eye gaze extracted from a mobile eye movement device (in-action).

Another challenging issue that arises in the domain of visual expertise is that the majority of eye-tracking-based studies do not implement in their reports information about how the various eye-tracking parameters are calculated, for example, which specific configurations in the eye movement event-detection algorithm to detect fixations and saccades were used. However, this is especially important as an increasing number of researchers are analyzing their gaze data with more advanced external analysis tools and scripts (Dolezalova & Popelka, 2016; Panetta et al., 2020) that are based on the raw eye-tracking data extracted from the eye tracker. We, therefore, aim to demonstrate the extent to which different configurations may affect the detection of fixations and saccades and thus also the (interpretation of) the results. The above outlined GRI is a suitable measure to investigate how different configurations affect the results of eye movement experiments, as it is a single-valued measure that allows a straightforward comparison.

## 1.1 Professional Vision and Visual Expertise of Teachers

Professional vision is commonly used as a conceptual framework in the field of cognitive-oriented teacher research (Goodwin, 1994; Seidel & Stürmer, 2014). The concept implies a two-step-process: (1) noticing, which describes teachers' ability to selectively direct attention to relevant events in the classroom; and (2) knowledge-based reasoning, which refers to teachers' ability to interpret these events based on their professional knowledge (Seidel & Stürmer, 2014; van Es & Sherin, 2010). Thereby, noticing and knowledge-based reasoning are not isolated processes, but interact with each other (Seidel & Stürmer, 2014). Teachers' professional knowledge drives teachers' attentional processes in a top-down process (i.e., selective attention inferred from their knowledge) and, in turn, noticing activates teachers' knowledge in form of curriculum scripts and classroom routines stored in their long-term memory (i.e., teachers can make sense of what they see) (Lachner et al., 2016). This implies that professional vision is formed primarily through consistent practice over many years in which teachers accumulate professional knowledge. It, therefore, indicates that professional vision is primarily a characteristic of experienced teachers (Berliner, 2001). Eye-tracking studies at the intersection of professional vision and visual expertise provide further evidence that teachers' visual processes change as their expertise increases (Grub et al., 2020; Kosel, Holzberger, & Seidel, 2021; van den Bogert et al., 2014). For example, it has been found that experienced teachers compared to novice teachers are able to distribute their attention more evenly across students (van den Bogert et al., 2014) and monitor a larger group of different students during teaching (Kosel, Holzberger, & Seidel, 2021). Beyond these findings especially relevant to classroom management, studies have shown that experienced teachers, similar to domain experts in other vision-intensive fields (Gegenfurtner et al., 2011; Gegenfurtner et al., 2022), have shorter but more fixations whereby domain novices have longer but fewer fixations. Since an important assumption is that fixations indicate that information is perceived and processed cognitively (Rayner, 2009), the results suggest that experts encode information more rapidly because of their more advanced and fine-graded knowledge structures that drive visual attention in a top-down process (Gegenfurtner et al., 2022).

In contrast, novices do not have this accumulated knowledge, and their attention is driven more by external and salient features of the visual stimulus in a bottom-up process (Gegenfurtner et al., 2022). The rapid information processing of experienced teachers (reflected in short fixation durations) is also consistent with Ericsson and Kintsch's (1995) theory of long-term working memory. They stated that experts increase the capacity of their working memory by building retrieval structures in their long-term memory (see also Gegenfurtner et al., 2022). The knowledge embedded in this retrieval structure is available in the working memory and enables experts to process visual information more rapidly compared to novices that have not yet fully developed a knowledge-based retrieval structure. In other words, relevant for visual expertise is not only a large amount of domain-relevant knowledge but also a superior organization of this knowledge. In addition, the ability of experts to process more information (reflected in a higher number of fixations) is related to the assumptions of Haider and Frensch's (1996) information reduction hypothesis. They argue that experts optimize the amount of information processed by separating task-relevant from task-irrelevant information. Ignoring redundant information leads to experts having more capacity in their working memory to process more relevant information. Both theories are also important parts of the Cognitive Theory of Visual Expertise (CTVE; Gegenfurtner et al., 2022) which covers further important aspects of visual expertise (e.g. parafoveally and holistic information processing). Taken together, the outlined assumptions help to understand experienced teachers' faster and more automated information processing involving less conscious effort, suggesting that experienced teachers encode and update dynamic teaching situations (with many and short fixations) more rapidly (Gegenfurtner et al., 2020; Grub et al., 2020).

To be able to capture visual expertise using a single and expertise-sensitive metric, Gegenfurtner and colleagues (2020) introduced the so-called Gaze Relation Index (GRI) in the field of visual expertise. The GRI is defined as the ratio of mean fixation duration to mean fixation count (Gegenfurtner et al., 2020). Based on the previous empirical studies (e.g., Wolff et al., 2016) and how the GRI is calculated, it can be inferred that the GRI should be higher for novice teachers than for experienced teachers. Since the GRI is still emerging in the field of visual expertise, the number of studies to date is limited. In the study by Gegenfurtner and colleagues (2020), the GRI was calculated for dynamic 3D medical visualizations. They found that the GRI was slightly, but statistically non-significant, higher for novices compared to experts (Gegenfurtner et al., 2020). In the educational context, Grub and colleagues (2022) analyzed the GRI in a standardized experimental design in which experienced and novice teachers perceived various classroom situations via short video sequences. Contrary to their hypothesis, they found no differences in the number and duration of fixations and thus no differences in GRI (Grub et al., 2022). However, as discussed by Gegenfurtner (2020), the full potential of the GRI might come to light when the experiment is situated outside the lab, using mobile eye-tracking "to mirror the full complexity of visual input that experts routinely deal with in their everyday work surroundings" (Gegenfurtner et al., 2020, p. 38). However, the number of studies that analyzed mobile eye-tracking data to explore expertise differences concerning the number and duration of fixations (the basis for the GRI) is limited (Huang et al., 2021). While the in-action study by Huang and colleagues (2021) confirms expected expertise differences regarding the two metrics, the findings of on-action eye-tracking studies are more heterogeneous (Grub et al., 2022; Kosel, Holzberger, & Seidel, 2021; van den Bogert et al., 2014; Wolff et al., 2016). One reason for this, as Gegenfurtner and colleagues (2020) described, could be that eye-tracking experiments in the laboratory cannot capture the full dynamic complexity that can only be recorded with mobile eye-tracking in teachers' natural work environment. Thus, there is a need to further explore novice and experienced teachers' visual expertise measured with the GRI using mobile eye-tracking data.

## 1.2 Classify eye movements using event-detection algorithms

Across all academic disciplines, eye-tracking-based studies rely on eye movement event-detection algorithms to analyze raw data and classify different types of eye movement, such as fixations (moments when the eye is relatively still and visual information is processed) and saccades (rapid eye movements between two or more phases of fixation). There exists a large number of different algorithms today (see for a review and evaluation of different algorithms: Andersson et al., 2017). Event-detection algorithms can be broadly grouped into dispersion- and velocity-based algorithms (Andersson et al., 2017). One of the most frequently used velocity-based algorithms for detecting fixations is the Identification by Velocity Threshold (I-VT). This algorithm uses only one parameter, the fixed velocity threshold for saccade detection where "fixations are segments of samples with point-to-point velocities below the set velocity threshold, and saccades are segments of the sample with velocities above this threshold" (Andersson et al., 2017, p. 618). The fixed and a-priori-defined velocity is most usually given in visual degrees per second (°/s). Commonly used values for the velocity threshold in lab-based eye-tracking studies range between 5 and 50°/s, using lower values for oculomotor studies and higher values for cognitive studies (Andersson et al., 2017). The I-VT algorithm is implemented in most of the recent commercial eye-tracking software like Tobii Pro (Tobii, 2022). However, since fixation is a fundamental parameter of most eye-tracking studies, outcomes depending not only on the used algorithm to separate fixations from saccades (Salvucci & Goldberg, 2000) but also on the different velocity thresholds employed for the algorithms (Andersson et al., 2017; Holmqvist et al., 2015). In other words, different velocity thresholds might produce significantly different results (Salvucci & Goldberg, 2000). The various velocity thresholds can be easily changed in most software solutions. In Tobii Pro, for example, velocity thresholds of 30°/s and 100°/s are pre-stored (30°/s = fixation filter; 100°/s attention filter). In this context, Hossain and Miléus (2016) compared different velocity thresholds for fixation identification for low-sample-rate mobile eye-trackers like the Tobii Pro Glasses 2. They point out that the IV-T fixation filter does not perform as well on mobile eye-tracker as it did on lab-based eye trackers—especially when a lot of head movements are involved in the recordings. The problem here is that head movements have an impact on velocities and many fixations would not be detected by the IV-T algorithm. They found that the default setting of 30°/s underestimates the periods during which a participant gathers information because a large proportion of smooth pursuits (eye movements in which the eyes remain fixated on a moving object) and vestibular-ocular reflex (VOR; stabilizing eye movements in the opposite direction of head movements) are classified as saccades. One way to counter this is to increase the velocity threshold on the mobile eye-tracker. Using the 100°/s attention filter would overestimate information gathering because fixations, smooth pursuits, VOR periods and 10-15% of short saccades will be classified as fixations (Hossain & Miléus, 2016). However, Hossain and Miléus (2016) found the highest precision for fixation detection in mobile eye-tracking using a velocity threshold between 90°/s and 100°/s, when head movements are involved and not compensated for with external gyroscope data. Overall, since some studies and technical reports point out that results significantly change with different velocity thresholds (Hossain & Miléus, 2016; Olsen, 2012; Salvucci & Goldberg, 2000), the selected velocity threshold must be set out in research studies to make results comparable. Most studies in the educational context (Chaudhuri et al., 2021; Cortina et al., 2015) and in other fields such as aviation (Weibel et al., 2012) where mobile eye tracking is used, however, do not specify the velocity thresholds used to detect fixations and saccades.

In addition to velocity thresholds, the so-called fixation merging is another configuration of the IV-T algorithm that needs to be addressed. The basic idea of merging fixations is that very short fixations (i.e. too short fixations do not reflect cognitive processing) are merged with the next longer fixation, which is in its vicinity (within 0.5° of visual angle) (Tobii, 2022). Merging can be set automatically in the Tobii software package (Olsen, 2012; Tobii, 2022). However, fixation merging has consequences for the classification of the number of fixations and thus for the results of fixation-based metrics such as the GRI. However, the extent of this effect has not yet been described, which makes an assessment together with different velocity thresholds relevant for future eye-tracking studies in the context of visual expertise.

**1.3 The present study**

In the present study, we aimed to explore teachers' visual expertise following an expert-novice paradigm. Established expertise theories and prior empirical findings point to the fact that teachers, through deliberate practice for a long period, develop visual expertise which leads to qualitatively enriched and superior ways of visually perceiving and processing information when compared to novices. Two of the expertise-sensitive eye-tracking metrics are the number of fixations and the average duration of these fixations. The introduced GRI is based on the relation between both parameters and can be used as a single value metric to assess visual expertise in vison-intense domains like teaching. However, until now the GRI is a seldom explored metric, and evidence is limited to lab-based on-action eye-tracking studies (Gegenfurtner et al., 2020; Grub et al., 2022). Therefore, the first aim was to use the GRI to measure teachers' visual expertise based on real-world gaze collected with a mobile eye-tracking device during instruction. The second aim of the present study was to investigate the impact of various velocity thresholds for eye movement identification using the Identification by Velocity Threshold (I-VT) algorithm and fixation merging on the eye-tracking parameter/GRI metric. This study is hypothesis-driven and involves two related research questions:

1. Is the gaze relational index (GRI) higher for experienced compared to novice teachers? We aimed to explore the potential utility of GRI as an indicator of visual expertise. Based on previous findings, we hypothesized that experienced teachers use more top-down knowledge-based processing of visual information, leading to their ability to scan the visual field more rapidly.

Thus, we expected more and faster fixations among experienced teachers. Novice teachers in comparison, use more button-up salient-based processing of visual information, resulting in fewer and longer fixations. Therefore, we expect the GRI to be higher for novice teachers than for experienced teachers.

2. How do the eye-movement parameters (fixations, duration of fixations) and the GRI change…

    a) depending on the choice of velocity thresholds for eye movement detection based on the Identification by Velocity Threshold (I-VT) algorithm?
    b) depending on fixation merging?

We expected that the different velocity thresholds would lead to different results regarding the detection of fixations and saccades, thus affecting the GRI. Based on the logic behind the IV-T algorithm, we expected that the lower the selected velocity threshold, the fewer eye movements were classified as fixations. However, derived from eye-tracking protocols and studies (Andersson et al., 2017; Olsen, 2012), we hypothesize that this is not a linear process, i.e., the velocity threshold of 30°/s compared to 60°/s does not classify half of the eye movements as fixations, mainly because using higher velocity thresholds more smooth pursuit eye movements and slow saccades were identified as fixations. Furthermore, we expected that fixation merging significantly reduces the number of fixations and therefore the GRI of a participant. The extent to which outcomes differ is difficult to predict, so this research question is exploratory in nature.

## 2 Methods

### 2.1 Participants

The data were obtained from four in-service mathematics teachers (two females, two males). Each teacher gave a lesson ranging between 60 - 90 min in four different higher secondary schools (grade 9) in Germany. All participating teachers taught similar content (matrix calculus) at the time of the data collection. In addition, the sampled lesson was minimally predetermined to allow for some consistency across teachers and their individual lessons - teachers were given 5min. of their class to recap the topic and tasks of the last lesson and the remaining time to introduce a new piece of content. Two of the participating teachers are novices with an average teaching experience of 1.5 years, while the other two teachers are experienced teachers with an average teaching experience of 11 years. Teachers were between 27 and 62 years old, (M=37.25, SD=16.64). Class sizes ranged from 14 to 24 students (69 students total).

**2.2 Procedure**

Mobile eye-tracking recording took place during a regular class period, chosen to interfere as little as possible with the regular lesson plan. We used a Tobii Pro Glasses 2 with a temporal resolution of 100 Hz to collect eye movement data (Tobii, 2022). Before the recordings started, a calibration of the eye-tracking glasses was performed until a satisfactory calibration was achieved. All participating teachers were advised not to move their eye-tracking glasses during the recording of eye movements. After the recording, the participating teachers were interviewed through a questionnaire (assessment of the lesson, demographic data, professional experience, etc.).

**2.3 Data (pre-)processing**

*2.3.1 Data Collection*

We exported the raw data using the Tobii Lab Analysis Software (Tobii, 2022), which gave us information about the eye and gaze positions at each recording timestamp, and performed all subsequent fixation calculations in python. For each timestamp, we stored the time since the beginning of the recording in milliseconds, the pupil positions of the left and right eye at this timestamp in 3D space, the gaze points at this timestamp in 3D space and a 2D representation of the gaze points at this timestamp. The time of recordings per participant varied between 24 and 68 minutes. To control for these time differences and to limit their impact on the eye-tracking parameter, we extracted for each person all eye movements of the first 20 minutes of the recording and discarded the rest of the data for our analysis.

*2.3.2 Fixation Classification Algorithm*

**Fixation calculation.** We based our fixation calculation on the Velocity-Threshold Identification (I-VT) algorithm, as described in (Salvucci & Goldberg, 2000) and (Olsen, 2012). First, we calculated the point-to-point velocities for each pair of consecutive recording timestamps (t1,t2), by performing the following steps:

1) We calculated the timestamp of the exact point in time between t1 and t2, by taking the mean of t1 and t2.
2) We calculated the position of the left eye at the timestamp t1t2 by taking the mean of the left eye position vector at t1 and the left eye position vector at t2. We did the same for the right eye.
3) We calculated the visual angle between the left eye position at timestamp t1t2, the gaze position at t1, as well as the gaze position at t2. We did the same for the right eye. This gave us an indicator of how far the gaze has moved from timestamp t1 to timestamp t2.
4) We divided the visual angle by the time between t1 and t2 in seconds. This gave us the angular velocity of an eye movement in degrees/second at timestamp t1t2.
5) We aggregated the velocity of the left and right eye by taking the mean of both velocities. If the velocity of one eye could not be inferred (e.g. because the person had blinked with one eye at time t1 or t2 or both), we took the velocity of the other eye. If velocities of both eyes could not be inferred, the sample was declared an invalid value.
6) We calculated the gaze positions in 2D and 3D space at timestamp t1t2, by taking the mean of each gaze position on time t1 and time t2.

7) For each point, we stored the timestamp t1t2, the angular velocity v_t1t2 at this timestamp, the gaze points at t1t2, and the eye position at t1t2.
8) Next, we labeled all points with a velocity below or equal to the velocity threshold parameter as fixations and all points above the threshold as a saccade. To study the impact of this velocity parameter on eye-tracking parameters, such as duration and number of fixations, we performed our analysis with different threshold values between 10 and 150 (stepsize=10). In addition, we put special attention to the velocity threshold of 30°/sec, as this is used per default in the fixation filter of the Tobii Lab Analysis Software, and the velocity threshold of 100°/sec, as this is used per default in the attention filter of the Tobii Lab Analysis Software (Tobii, 2022).

**Building fixation groups.** After this, we build fixation and saccade groups by merging all consecutive points containing a fixation to a fixation group, all consecutive saccades to a saccade group, and all consecutive points with an invalid value to an invalid group. For each group, we defined the start time as the point of time between the timestamp of the first sample in this group and the timestamp of the last sample of the preceding group. Similarly, we defined the end time as the point of time between the timestamp of the last sample in this group and the timestamp of the first sample of the preceding group. We calculated the duration of the group by subtracting the start time from the end time and calculating the eye and gaze positions by taking the mean of all eye and gaze points in this group. We furthermore stored the eye movement type (fixation, saccade, invalid) as well as a counter for fixations, saccades, and invalids.

**Fixation merging.** We then merged fixation groups, which were divided by a saccade or invalid value, but were close in time and space. We did this using the following steps

1) For each pair of subsequent fixation groups f1,f2, we calculated the time between the end of f1 and the beginning of f2. If this time was shorter than a threshold (max_time_betw_fixations), we continued with step 2, otherwise, we continued with the next fixation pair. We used a max_time_betw_fixations threshold of 75 milliseconds as recommended in (Olsen, 2012).

2) We calculated the visual angle between f1 and f2 by using the mean eye position of f1 and f2, the gaze position in f1, and the gaze position in f2 for the left eye. We did the same for the right eye and merged the visual angles of both eyes as described in step 5 in fixation calculation. If the overall angle was shorter than a threshold (max_angle_betw_fixations), we merged the fixation groups in the same way as merging consecutive fixations. All saccades and invalid values between f1 and f2 were thus discarded. We used a max_angle_bw_fixations threshold of 0.5 degrees, as recommended in (Olsen, 2012).

To study the impact of fixation merging on eye-tracking parameters, we performed our analysis once with and once without fixation merging.

**Eye-tracking parameter calculation.** We then continued to examine individual eye-tracking parameters. For each person, we calculated the number of fixations $fixnr_{Person\ x}$ as well as the mean fixation duration $fixdur_{Person\ x}$, meaning the sum of lengths of all fixations of this person, divided by the number of fixations. Furthermore, we defined the Gaze Relational Index of a person as $GRI_{Person\ x} = fixnr_{Person\ x}/fixdur_{Person\ x}$ and calculated this index for each person.

For an expert-novice contrast, we calculated the mean fixation number $fixnr_{Group\ x}$ by taking the mean of the fixation numbers of all participants in this group. Furthermore, we calculated the mean fixation duration of this group $fixdur_{Group\ x}$, by taking the mean of the mean fixation durations of all participants in this group. We then calculated the Gaze Relational Index of a group, as defined in (Gegenfurtner et al., 2020), by using the following formula: $GRI_{Group\ x} = fixnr_{Group\ x}/fixdur_{Group\ x}$.

## 3 Results

### 3.1 Differences between experienced and novice teachers' gaze relational index

The first research question examined the extent to which experienced and novice teachers differ in the GRI. Table 1 shows the eye movement parameters number of fixations, duration of fixations, and GRI using the velocity threshold of 30°/s and 100°/s, separated by expertise level. Descriptive results indicate that experienced teachers had more fixations, shorter fixation durations, and a lower GRI compared to novice teachers. Although the trend of the results has not changed, varying velocity thresholds have an impact on the eye movement parameter/GRI. For example, while the difference between expert groups in GRI is marginal at a velocity threshold of 30°/s, novice teachers' GRI is more than double that of experienced teachers at a velocity threshold of 100°/s.

**Table 1**

*Group-based eye-tracking parameter and GRI with velocity threshold of 30/100 and no merging of fixations*

|  | Fixation Number | | Mean Fixation Duration | | GRI |
|---|---|---|---|---|---|
|  | *M* | *SE* | *M* | *SE* |  |
| Experts (VT 30°/s) | 4319.50 | 566.50 | 125.52 | 17.82 | 0.030 |
| Novices (VT 30°/s) | 3991.50 | 569.50 | 156.35 | 3.35 | 0.039 |
| Experts (VT100°/s) | 3981.50 | 674.50 | 186.79 | 29.51 | 0.047 |
| Novices (VT100°/s) | 2569.50 | 401.50 | 326.37 | 18.70 | 0.127 |

**3.2 Impact of varying velocity threshold on the number of fixations, duration of fixations, and the GRI**

The second research question examined the impact of varying velocity thresholds and fixation merging (yes/no) on eye movement parameters/GRI. Figure 1 shows the eye movement parameter/GRI for each person with each analysis. Merging categories seem to have little to no effect on the mean fixation duration, fixation number, and GRI of a person. However, as already indicated in the results of RQ1, the velocity threshold seems to have a high influence on the eye movement parameter/GRI: A higher velocity threshold leads to more samples being classified as fixation. As consecutive samples containing a fixation are merged, a higher velocity threshold implies a higher mean fixation duration. At the same time, a higher velocity threshold leads to a lower number of fixations for thresholds above 30-40°/s. At first sight, this seems to be counterintuitive, but is based on the merging of consecutive samples: For example, there could be 3 samples in the dataset, s1, s2, and s3. With a velocity threshold of 30, these would be classified as s1=fixation, s2=saccade, and s3=fixation, resulting in a fixation number of two. With a velocity threshold of 100 in contrast, s2 could be classified as fixation as well. As consecutive fixations are merged, this would result in a fixation number of one, meaning the fixation number decreases with an increasing velocity threshold. However, the above-identified relation is not linear, which means that the order of the participants in terms of their level of GRI is changed. For example, E1 has a higher GRI than N2 when using a velocity threshold of 30°/s, but a lower GRI than N2 when using a velocity threshold of 100°/s.

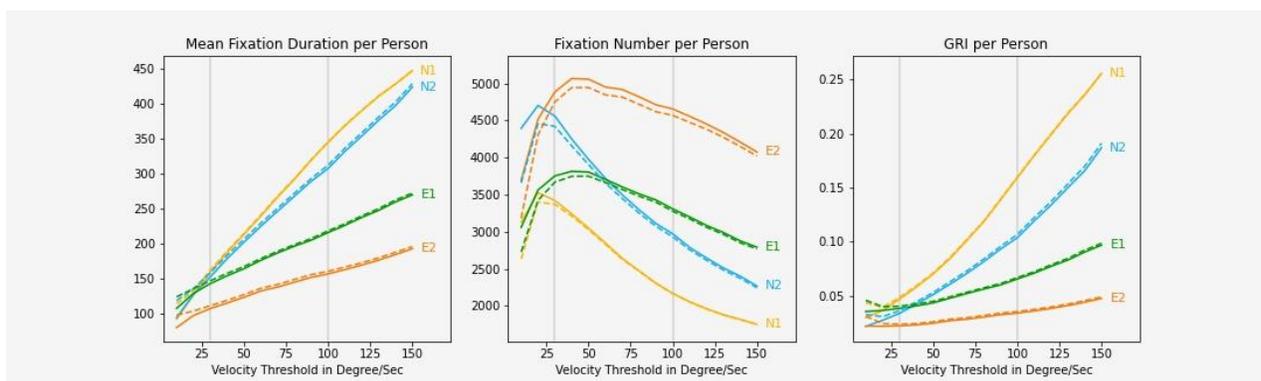

**Figure 1:** *Mean fixation duration, fixation number, and GRI per person for a fixation calculation with different velocity thresholds between 10 and 150 degrees per second. Solid lines represent the analysis without merging of fixation groups and dashed lines the analysis with fixation group*

*merging. The velocity thresholds of 30 and 100 are marked with a grey line. E1/2 = Experienced teachers, N1/2 = Novice teachers*

# 4 Discussion

The teaching profession heavily depends on visual information—teachers visually perceive, collect, and process information in a complex and dynamic classroom environment (Wolff et al., 2016). Over the last years, cognitively-oriented educational research found that experienced teachers develop domain-specific visual expertise which has not yet developed in novice teachers (Kosel, Holzberger, & Seidel, 2021; van den Bogert et al., 2014; Wolff et al., 2016). The present study aimed to contrast the visual expertise of experienced and novice teachers, as measured by the Gaze Relational Index (GRI), in highly dynamic real-world classroom environments using mobile eye-tracking data. Furthermore, the study explores how different configurations (varying velocity thresholds & fixation merging) of the IV-T algorithm for eye-movement classification affect the results of the study. In general, our findings correspond to the perceptual superiority of domain experts (indicated by a lower GRI) and findings suggest that different velocity thresholds for eye-movement identification significantly affected the results of our study.

## 4.1 Gaze Relational Index as an expertise-sensitive metric in research about teachers' visual expertise

We expected experienced teachers to process visual information faster and with more numerous fixations (characterizing the domain-specific superiority of experienced teachers in visual processing; Gegenfurtner et al., 2011), and thus need less time and effort to comprehend the complexity of classroom situations (Gegenfurtner et al., 2022). Therefore, the GRI (ratio of the mean number of fixations to the mean duration of fixations) was expected to be lower for experienced teachers than for novices. We were able to provide support for this hypothesis as we found experienced teachers have more fixations with shorter average fixation duration than novice teachers and thus, have a lower GRI compared to novice teachers. Contrary to other studies in the context of visual expertise of medical experts and novices (Gegenfurtner et al., 2020) as well as experienced and novice teachers (Grub et al., 2022), the calculated GRI in this study was more sensitive to expertise. One decisive reason for the heterogeneous results may play a role (besides more technical reasons, which will be discussed later): Compared to the outlined studies above, we have begun to step outside artificial classroom environments of laboratory setups toward the more natural conditions teachers usually face in real classrooms using mobile eye-tracking. It has been shown that eye movements in the real world generally vary more among participants (Dowiasch et al., 2020). Dowiasch and colleagues (2020) argue that this could be since mobile eye-tracking gaze recordings are generally much less restrictive than laboratory gaze recordings, allowing participants to behave more naturally. In this context, teachers often experience a much higher level of complexity in their real work environment, which is difficult to mirror in laboratory eye-tracking research. Therefore, a general transferability of results from eye-movement measurements in the laboratory to the real world seems difficult, although researchers underline to better understand visual behavior/expertise in natural environments (Dowiasch et al., 2020; Gegenfurtner et al., 2020). We have taken this step with this study and can confirm assumptions on expertise differences in the GRI. Results might indicate that experienced teachers` superior visual processing comes to the surface, especially in complex and dynamic real-life situations.

**4.2 Varying velocity thresholds for eye-movement identification influence the GRI**

Across all research areas, eye-tracking-based studies face the critical challenge of transforming the raw gaze signals of the eye-tracker (i.e., gaze origin and the gaze direction) into meaningful gaze parameters (i.e., fixations, saccades) (Olsen, 2012; Tobii, 2022). Not only that there are numerous algorithms for this task available, the algorithms often work with different and customizable configurations (Andersson et al., 2017). We investigated the impact of different velocity threshold settings for one of the most commonly used algorithms (IV-T: Andersson et al., 2017; Olsen, 2012) on the results of our mobile eye-tracking study (RQ2a). The results indicate that the choice of a velocity threshold influences the mean fixation duration and fixation number per person and it consequently influences the GRI per person. In addition, we identified that the selection of the velocity threshold not only influences the absolute size of the GRI but also the rank order of participants regarding their GRI. In other words, the different velocity thresholds do not have a linear effect on the number and duration of fixations and the GRI. Concerning the default fixation filter (30°/s) and attention filter (100°/s) provided by Tobii (Olsen, 2012; Tobii, 2022), the results are less influenced when interpreted on the averaged group level (experts vs. novices) than on the individual level (e.g., comparison of single participants). However, because eye-tracking studies have comparatively few participants compared to other traditional study designs (i.e., questionnaire surveys), the presumed influence on the results is all the more striking (for example, when comparing group means). From this more methodological perspective, we argue that the upcoming heterogeneity that occurs in the results of visual expertise studies (e.g., described by Klostermann & Moeinirad, 2020) regarding the number and duration of fixations of domain experts may be due not only to different study contexts (e.g., varying professional domains or tasks) but also to the choice of a specific velocity threshold. Thereto, the choice of velocity threshold should be regularly reported in publications. Furthermore, the process of fixation merging (RQ2b) did not affect our results compared to different velocity thresholds. This is probably due to the fact that in our data very few fixation groups are merged. Choosing higher parameters for the maximal time between fixations and the maximum angle between fixations would result in more fixations being merged and could consequently lead to higher differences between analyses. Since the manual settings of fixation merging are more restricted (in comparison to velocity thresholds) in current software packages (Olsen, 2012; Tobii, 2022), we assume that the influence concerning fixation merging in studies is reduced since default values are often maintained.

In sum, our study demonstrates the importance of transparently specifying configurations of algorithms for eye-movement classification in eye-tracking studies that base their interpretation on fixations and saccades. This is one step to valid, reliable, and objective measurements of eye movements in the field of visual expertise. Based on our results and in agreement with Hossain and colleagues (2016), we recommend using the 100°/s fixation filter when mobile eye tracking is used and head movements are involved.

5.2 Limitations and future directions

The present study has three main limitations that can be addressed by future research.

First, our study is limited to descriptive (group) comparisons mainly due to the small sample size. As the present results are exploratory, further research is needed to confirm these observed differences using a larger sample size.

Second, we have limited knowledge of the extent to which the GRI is related to the specific situations teachers face in the classroom. Therefore, the following considerations on this aspect must be taken into account: Our analysis showed that experienced and novice teachers differed in their visual behavior as measured by the GRI, but we know little about *how* they differed in their interpretation of what they saw. Future research should focus on a more comprehensive combination of eye-tracking and think-aloud protocols to understand how the GRI relates to the underlying instructional situations that the teacher cognitively and visually faced during the eye-tracking recording. Another way to achieve this would be to code the first-person video recorded by the mobile eye-tracking afterward to investigate the GRI, for example, in different forms of instructional interaction (frontal instruction, group work). In this context, it should also be noted that we performed a gaze-based approach (analyses are based only on the fixations) and have not integrated any Areas of Interest (AOIs)—this means that we did not take into account the distribution of attention to specific areas in the classroom. This brings up an important point that should be considered in future research. To realize the full potential of the GRI, future studies should integrate relevant AOIs in mobile eye-tracking data to analyze which areas in the classroom are being processed with a high or low GRI (for example, to analyze the GRI in relation to task-related and task-irrelevant areas). To summarize this above-described outlook, there is a further need to understand in which situations the visual expertise of experienced teachers come to the fore.

Third, we have focused on two essential parameters (velocity threshold, fixation merging) for the various configurations of the IV-T algorithm and ignored other aspects such as interpolation (filling gaps in raw eye-tracking data in which no signal was recorded) or active noise reduction (e.g., noise may be caused by imperfect system settings) (Tobii, 2022). The extent to which these (often manufacturer-specific) methods of data preprocessing influence the results (especially when studies use different eye-tracking devices) is to be clarified.

A final comment to the GRI is warranted. It should be noted that the GRI depends on the length of the recording: While the mean fixation duration should remain relatively constant over a period of time, the number of fixations increases with each recording minute, leading to a decrease in the GRI value. Therefore, we recommend analyzing participants always over the same amount of time and to state the recording time when reporting a study.

5.3 Conclusion

Our study leads to two important consequences concerning research on teachers' visual expertise: The GRI might serve as a sensitive measure of visual expertise when using mobile eye-tracking data. The lower GRI of experienced teachers indicates that they have a distinct visual behavior, which is indicative of their fine-grained domain-specific knowledge organization that is reflected in their visual expertise. Regardless of selected velocity thresholds for identifying fixations, the experienced teachers showed shorter and more fixations that resulted in a lower GRI. However, from a methodological viewpoint, the study also showed that the selected velocity threshold parameters alter and, in the worst case, bias the results of an eye-tracking study. Therefore, in the interest of further generalizability of the results within visual expertise research, researchers are encouraged to be transparent about reporting their configurations relevant to eye movement identification.